\documentstyle[preprint,aps,
]{revtex}
\title {Electromagnetically-induced transparency and light storing of a pair of pulses }
\author{A. Raczy\'nski$^{1}$
\footnote{email: raczyn@phys.uni.torun.pl}, J. Zaremba$^{1}$ and
S. Zieli\'nska-Kaniasty$^{2}$}
\address{$^{1}$Instytut Fizyki, Uniwersytet Miko\l aja Kopernika,
       ul.Grudzi\c{a}dzka 5,
       87-100 Toru\'n, Poland,\\$^{2}$ Instytut Matematyki i Fizyki,
       Akademia Techniczo-Rolnicza, Al. Prof. S. Kaliskiego 7, 85-796
       Bydgoszcz, Poland.}
\begin{document}

\maketitle

\begin{abstract}
Electromagnetically-induced transparency and light storing are
studied in the case of a medium of atoms in a double $\Lambda$
configuration, both in terms of dark- and bright-state polatitons
and atomic susceptibility. It is proven that the medium can be
made transparent simultaneously for two pulses following their
self-adjusting so that a condition for an adiabatic evolution has
become fulfilled. Analytic formulas are given for the shapes and
phases of the transmitted/stored pulses. The level of transparency
can be regulated by adjusting the heights and phases of the
control fields.
\end{abstract}
\newpage
\section{Introduction}
It is well known that an atomic medium irradiated by a control
laser field may become transparent for a signal field which in the
absence of the first field would be almost immediately absorbed.
This phenomenon, known as an electromagnetically-induced
transparency (EIT) \cite{a1,a2}, has recently been used to
drastically change the velocity of a light pulse or even to stop
or store it: by changing the control field in time it is possible
to make the medium opaque at the moment at which the signal pulse
is inside \cite{a3,a4,a5}. The pulse is then transformed into an
atomic coherent excitation which is rather robust against
relaxation and after quite a long time in the atomic scale it is
possible to switch the control field on again and to release the
trapped signal. Such processes have been observed experimentally
and explained theoretically both in the language of so-called
polaritons, being collective atom+field excitations, and in terms
of atomic susceptibily \cite{a6,a7}. An elementary atomic systems
for which such processes are possible is an atom with three active
(resonantly coupled) states in the $\Lambda$ configuration.

Adding a fourth active state and a second control field i.e.
extending the atomic system to a double $\Lambda$ configuration
allows one to consider new nonlinear, resonantly enhanced optical
processes \cite{a8}. In particular it is possible to
simultaneously propagate two optical pulses of different
frequencies through a medium in the conditions of EIT. In our
earlier paper \cite{a9} we have pointed out that one can stop one
pulse and release a pulse of a different frequency or two
different pulses. We have also mentioned that making the medium
transparent simultaneously for two pulses is possible provided
that the fields are in some special relation, which allows for an
adiabatic evolution of the system. If on the other hand the
initial conditions do not satisfy this relation the pulses are
partially absorbed, which is connected with nonadiabatic
phemomena. In the present work we give a detailed quantitative
analysis of the conditions of a joint medium transparency for two
pulses. A discussion in terms of dark- and bright-state
polaritons, presented in the next section, includes simple
analytic formulas which allow one to predict the amplitudes,
phases and time evolution of the transmitted and restored signals
and/or the value and space distribution of the atomic coherence
due to the trapped pulses. Section III shows how the
non-adiabaticity of the evolution is reflected on the atomic
susceptibility. Section IV contains a comparison of the
predictions of the polariton approach with the results of complete
numerical solutions of the Bloch-Maxwell equations.

\section{Polaritons}
We consider a medium composed of atoms in a double $\Lambda$
configuration presented in Fig. \ref{fig1}. Two driving fields
$\epsilon_{2,4}(t)$, induce transparency for two weak signal
fields $\epsilon_{1,3}$. The evolution of the atomic density
matrix $\sigma$ and the propagation of the signal pulses are
described by the a set of Bloch-Maxwell equations, while
propagation effects for the driving fields are neglected.

In the rotating-wave and slowly-varying-envelope approximations,
in the resonance conditions without relaxations for the
transitions shown in Fig. \ref{fig1}, we are left in the first
order with respect to the signal fields with the equations
\begin{eqnarray}
&&(\frac{\partial}{\partial t}+c\frac{\partial}{\partial
 z})R_{1}=S_{1},\nonumber\\
&&(\frac{\partial}{\partial t}+c\frac{\partial}{\partial
z})R_{3}=S_{3},\nonumber\\ &&\frac{\partial}{\partial
t}S_{1}=-\kappa_{1}^{2}(R_{1}+U_{2}\sigma_{bc}),\\
&&\frac{\partial}{\partial
t}S_{3}=-\kappa_{3}^{2}(R_{3}+U_{4}\sigma_{bc}),\nonumber\\
&&\frac{\partial}{\partial
t}\sigma_{bc}=U_{2}^{*}S_{1}+U_{4}^{*}S_{3}\nonumber.
\end{eqnarray}
In the above formulas
$\kappa_{1,3}^{2}=\frac{|d_{1,3}|^{2}\omega_{1,3}N}{4\epsilon_{0}\hbar}$,
$\omega_{j}$ is the frequency of the field $j$, $N$ is the density
of atoms, $\epsilon_{0}$ is the vacuum electric permittivity,
$d_{j}$ are the transition matrix elements: $d_{1}=d_{ab}$,
$d_{2}=d_{ac}$, $d_{3}=d_{db}$, $d_{4}=d_{dc}$,
$R_{j}=\frac{\epsilon_{j}d_{j}^{*}}{2\hbar \kappa_{j}}$, $U_{2,4}=
\frac{\epsilon_{2,4}d_{2,4}^{*}}{2 \hbar \kappa_{1,3}}$,
$S_{1}=-i\kappa_{1} \sigma_{ba}$, $S_{3}=-i\kappa_{3}\sigma_{bd}$
and $\sigma=\sigma(z,t)$ is the atomic density matrix after
transforming-off the rapidly oscillating terms.

The adiabatic approximation would consist in setting the time
derivatives of $S_{1,3}$ equal to zero, which would mean an
evolution during which the atomic upper states are not populated
at all. However, one can see from the third and fourth equation in
the set (1) that this is possible if the four laser fields remain
in a certain proportion:
$\frac{R_{1}}{U_{2}}=\frac{R_{3}}{U_{4}}$. This reflects the fact
that during an adiabatic evolution each atom makes a continuous
transition from the initial state $b$ to the dark state, which is
a superposition of the two lower states $b$ and $c$ with the
coefficients determined by the instantaneous values of the
intensities of the signal and control fields. However, in a double
$\Lambda$ system this combination is a dark state simultaneously
for the two $\Lambda's$ only in the special situation.

After eliminating the variables $S_{1,3}$ the above equations can
be rewritten in a new set of variables - the so-called polaritons
\begin{eqnarray}
&&R_{1}=\exp(i \arg(U_{2}))(\cos\theta\cos\phi
\Psi+\sin\theta\cos\phi\Phi+\sin\phi X),\nonumber\\ &&R_{3}=\exp(i
\arg(U_{4}))(\cos\theta\sin\phi
\Psi+\sin\theta\sin\phi\Phi-\cos\phi X),\\
&&\sigma_{bc}=-\sin\theta \Psi+\cos\theta\Phi,\nonumber
\end{eqnarray}
where $\sin\theta=(|U_{2}|^{2}+|U_{4}|^{2}+1)^{-1/2}$ and
$\tan\phi=\frac{|U_{4}|}{|U_{2}|}$.

The evolution equations for the polaritons in the case of
time-independent driving fields read
\begin{eqnarray}
&&\frac{\partial}{\partial t}\Psi+c \cos^{2}\theta
\frac{\partial}{\partial z} \Psi+c\sin\theta\cos\theta
\frac{\partial}{\partial z} \Phi=0,\nonumber\\
&&\frac{-\sin^{2}\theta}{\cos\theta}\frac{\partial^{2}}{\partial
t^{2}} \Psi+\sin\theta \frac{\partial^{2}}{\partial t^{2}} \Phi=
-\frac{1}{\sin\theta}(\kappa_{1}^{2}\cos^{2}
\phi+\kappa_{3}^{2}\sin^{2}\phi)\Phi-\sin\phi\cos\phi(\kappa_{1}^{2}
-\kappa_{3}^{2})X,\\ &&\frac{\partial}{\partial
t}(\frac{\partial}{\partial t}+c\frac{\partial}{\partial
z})X=-\frac{\sin\phi\cos\phi}{\sin\theta}(\kappa_{1}^{2}-\kappa_{3}^{2})
\Phi-(\kappa_{1}^{2}\sin^{2}\phi+\kappa_{3}^{2}\cos^{2}\phi)
X.\nonumber
\end{eqnarray}
In the case of a single $\Lambda$ system (i.e. for $X=0$) an
adiabatic evolution meant that the dark state polariton traveled
with the velocity $c\cos^{2}\theta$ keeping its shape, with the
bright state polariton $\Phi=0$. The only necessary condition of
adiabaticity was that the pulses 1 and 2 should be smooth enough.
For a double $\Lambda$ system, again assuming smoothness of all
the pulses, the new element is that the bright state polariton $X$
is in general different from zero at the beginning of propagation:
the decomposition of the incoming signal fields is given by Eqs
(2) (with $\Phi=0$)
\begin{eqnarray}
&&R_{1}^{0}=\exp[i\arg(U_{2})_{0}](\cos\theta_{0}\cos\phi_{0}\Psi+\sin\phi_{0}
X),\nonumber\\
&&R_{3}^{0}=\exp[i\arg(U_{4})_{0}](\cos\theta_{0}\sin\phi_{0}
\Psi-\cos\phi_{0} X).
\end{eqnarray}
Then during the evolution $\Psi$ keeps its shape while $X$ is
damped. Thus the conditions of adiabaticity are gradually created
and the three dynamical variables tend to the following values
\begin{eqnarray}
&&R_{1}\rightarrow
\exp[i\arg(U_{2})]\frac{\cos\theta\cos\phi}{\cos\theta_{0}}
(\cos\phi_{0}\exp[-i\arg(U_{2})_{0}]R_{1}^{0}+\sin\phi_{0}\exp[-i\arg(U_{4})_{0}]
R_{3}^{0})\nonumber\\ &&R_{3}\rightarrow
\exp[i\arg(U_{4})]\frac{\cos\theta\sin\phi}{\cos\theta_{0}}
(\cos\phi_{0}\exp[-i\arg(U_{2})_{0}]R_{1}^{0}+\sin\phi_{0}\exp[-i\arg(U_{4})_{0}]
R_{3}^{0}),\\
&&\sigma_{bc}\rightarrow-\frac{\sin\theta}{\cos\theta_{0}}
(\cos\phi_{0}\exp[-i\arg(U_{2})_{0}]R_{1}^{0}+\sin\phi_{0}\exp[-i\arg(U_{4})_{0}]
R_{3}^{0}).\nonumber
\end{eqnarray}
The argument of the functions $R_{1,3}^{0}$ in the r.h.s of Eqs
(5) has to take into account the shift by $\int_{0}^{t} c
\cos^{2}\theta(\tau)d\tau$ during the time $t$. In particular if
we assume that at $t=0$ the edge of the pulse reaches the sample,
when calculating the position of the maximum we have to take into
account that the maximum moves first with the velocity $c$ until
it reaches the sample and later travels with the velocity
$c\cos^{2}\theta$.

The above formulas allow one to predict the shape and the
numerical parameters of the transmitted pulses in typical EIT,
when the control fields are kept constant, as well as in the
process of light storing, when the control pulses are slowly
varying.  Note in particular that a single initial signal in the
presence of two control fields leads to an appearance of the other
signal pulse. In Section IV we will compare the predictions based
on these formulas with the results of the numerical solutions of
the Bloch-Maxwell equations.

\section{Atomic susceptibility}
A complementary picture of the above evolution is obtained in
terms of the susceptibility. If we rewrite the first three
equations of Eq. (1) in the original variables admitting the
detuning and relaxations terms
\begin{eqnarray}
&&i\dot{\sigma}_{ba}=\frac{1}{2\hbar}\epsilon_{1}d_{1}^{*}+\Omega_{2}\sigma_{bc}+
\Delta_{1}\sigma_{ba}\nonumber\\
&&i\dot{\sigma}_{bd}=\frac{1}{2\hbar}\epsilon_{3}d_{3}^{*}+\Omega_{4}\sigma_{bc}+
\Delta_{3}\sigma_{bd}\\
&&i\dot{\sigma}_{bc}=\Omega_{2}\sigma_{ba}+\Omega_{4}\sigma_{bd}+
\delta\sigma_{bc},\nonumber
\end{eqnarray}
where $\Omega_{2,4}=\frac{\epsilon_{2,4}d^{*}_{2,4}}{2\hbar}$,
$\hbar\Delta_{1}=E_{b}
-E_{a}-\hbar\omega_{1}-\frac{i}{2}\Gamma^{a}$,
$\hbar\Delta_{3}=E_{d}
-E_{b}-\hbar\omega_{3}-\frac{i}{2}\Gamma^{d}$ and
$\hbar\delta=E_{b}+\hbar\omega_{1}-E_{c}- \hbar\omega_{2}-i\gamma$
are the detunings including the relaxation rates for the
coherences $ab$, $db$ and $bc$.

If we now pass to the frequency domain, assuming that
$\omega_{1}-\omega_{2}=\omega_{3}-\omega_{4}$, we can calculate
the elements of the density density matrix $\sigma$ and express
the components of the polarization in terms of the signal fields
\begin{eqnarray}
&&Nd_{1}\sigma_{ba}(\omega)=2\pi\epsilon_{0}[\chi_{11}(\omega)\epsilon_{1}
(\omega)+ \chi_{13}(\omega)\epsilon_{3}(\omega)]\nonumber\\
&&Nd_{3}\sigma_{bd}(\omega)=2\pi\epsilon_{0}[\chi_{31}(\omega)\epsilon_{1}
(\omega)+ \chi_{33}(\omega)\epsilon_{3}(\omega)],
\end{eqnarray}
where
\begin{eqnarray}
\chi_{11}(\omega)=\frac{N|d_{1}|^{2}}{4\pi\hbar\epsilon_{0}}
\frac{-(\Delta_{3}-\omega)(\delta-\omega)-
|\Omega_{4}|^{2}}
{(\Delta_{1}-\omega)(\Delta_{3}
-\omega)(\delta-\omega)-(\Delta_{1}-\omega)|\Omega_{4}|^{2}-(\Delta_{3}
-\omega)|\Omega_{2}|^{2}},\nonumber\\
\chi_{13}(\omega)=\frac{N d_{1}d_{3}^{*}}{4\pi\hbar\epsilon_{0}}
\frac{-\Omega_{2}\Omega_{4}^{*}}
{(\Delta_{1}-\omega)(\Delta_{3}
-\omega)(\delta-\omega)-(\Delta_{1}-\omega)|\Omega_{4}|^{2}-(\Delta_{3}
-\omega)|\Omega_{2}|^{2}},\nonumber\\
\chi_{31}(\omega)=\frac{N d_{1}^{*}d_{3}}{4\pi\hbar\epsilon_{0}}
\frac{-\Omega_{2}^{*}\Omega_{4}}
{(\Delta_{1}-\omega)(\Delta_{3}
-\omega)(\delta-\omega)-(\Delta_{1}-\omega)|\Omega_{4}|^{2}-(\Delta_{3}
-\omega)|\Omega_{2}|^{2}},\\
\chi_{33}(\omega)=\frac{N|d_{3}|^{2}}{4\pi\hbar\epsilon_{0}}
\frac{-(\Delta_{1}-\omega)(\delta-\omega)-
|\Omega_{2}|^{2}}
{(\Delta_{1}-\omega)(\Delta_{3}
-\omega)(\delta-\omega)-(\Delta_{1}-\omega)|\Omega_{4}|^{2}-(\Delta_{3}
-\omega)|\Omega_{2}|^{2}}.\nonumber
\end{eqnarray}
It is important to notice that under the assumptions of the
polariton analysis, i.e. in the resonance conditions and
neglecting the relaxations, the expressions for the susceptibility
$\chi$ become singular at $\omega=0$, e.g.,
\begin{eqnarray}
&&\chi_{11}(\omega)=\frac{N|d_{1}|^{2}}{4\pi\hbar\epsilon_{0}}
\frac{\omega^{2}-|\Omega_{4}|^{2}}{\omega^{3}-\omega(|\Omega_{2}|^{2}+
|\Omega_{4}|^{2})},\nonumber\\
&&\chi_{13}(\omega)=\frac{Nd_{1}d_{3}^{*}}{4\pi\hbar\epsilon_{0}}
\frac{\Omega_{2}\Omega_{4}^{*}}{\omega^{3}-\omega(|\Omega_{2}|^{2}+
|\Omega_{4}|^{2})},
\end{eqnarray}
with analogous expressions for $\chi_{31}$ and $\chi_{33}$. One
cannot thus speak of a transparency window and the pulses'
propagation occurs with a significant distortion. Fulfilling the
adiabaticity condition $\frac{R_{1}}{U_{2}}=\frac{R_{3}}{U_{4}}$
(cf. the third and fourth of Eqs (1)) means that the singularities
in Eqs (7) cancel out and the induced transparency inside a
transparency window of a finite size is possible. In that case
after substituting
$\epsilon_{3}=\epsilon_{1}\frac{d_{1}^{*}\Omega_{4}}{d_{3}^{*}\Omega_{2}}$
we obtain
\begin{equation}
Nd_{1}\sigma_{ba}=2\pi\epsilon_{0}\chi_{11}(\omega)\epsilon_{1}(\omega),
\end{equation}
with
\begin{equation}
\chi_{11}(\omega)=\frac{N|d_{1}|^{2}}{4\pi\hbar\epsilon_{0}}
\frac{-\omega}{\omega^{2}-(|\Omega_{2}|^{2}+|\Omega_{4}|^{2})},
\end{equation}
which is the expression as for a single $\Lambda$ system, with the
only difference that the denominator is corrected by the
$|\Omega_{4}|^{2}$term, which means that the transparency window
is widened compared with the case of a single $\Lambda$ system.

\section{Numerical illustration}
A numerical illustration of the above results is presented below.
We have performed calculations for the atomic model with
$E_{a}$=-0.10 a.u., $E_{b}$=-0.20 a.u, $E_{c} $=-0.18 a.u.,
$E_{d}$=-0.05 a.u., the width of the upper levels $a$ and $d$
connected with the spontaneous transitions to the lower levels $b$
and $c$ were taken $\Gamma_{b,c}^{a,d}=2.4\times 10^{-9}$ a.u.,
from which the dipole moments have been calculated; the relaxation
of the lower-states' coherence has been neglected. The medium
density has been taken $N=3\times 10^{-13}$ a.u. ($2\times
10^{12}$ cm$^{-3}$). The length of the sample was $10^{7}$ a.u
(0.5 mm). The initial length of the signal pulses was $10^{11}$
a.u. (2.4 $\mu$s) and their amplitudes, modeled by a sine-square,
were of order of $10^{-10}$ a.u., which correspond to the power
density of $3.5\times 10^{-4}$ Wcm$^{-2}$. The amplitudes of the
control pulses were larger by about an order of magnitude.

Because there are essentially two reasons of nonadiabaticity of
the evolution, namely the discontinuity of pulses or the fields'
failing to satisfy the proportion following Eq. (1), we first
check the role of the former effect. In Fig. \ref{fig2} we show
the shape of the initially rectangular pulse in a single $\Lambda$
system in a control field with a time-independent amplitude. The
evolution is initially nonadiabatic. However, the signal
propagating in the medium is gradually smoothed, which corresponds
to an absorption of the polariton $\Phi$. An analogous calculation
with a smooth (sine square) pulse yields to a very good
approximation a conservation of the pulse shape during the
propagation, which means an absence of the polariton $\Phi$ from
the very beginning. Because in our further investigations on
double $\Lambda$ systems we use sine square pulses, it follows
that we may indeed neglect the polariton $\Phi$, as mentioned in
Section II.

In Fig. \ref{fig3} we show the transmitted pulses 1 and 3 compared
with their initial shapes. The horizontal lines show the values
calculated from Eq. (5). The predictions of the heights of the
pulses are excellent. We have checked that Eq. (5) predicts
properly also the phases of the transmitted signal pulses.

The effective transparency for the two pulses can be regulated by
choosing the heights and phases of the control fields. In Fig.
\ref{fig4} we present the transmitted pulses 1 and 3 for two
different phase relations. By a sutiable choice of the phase of
any of the control fields one can regulate the heights of the
signal pulses.

If we switch the control fields off the medium becomes opaque for
the two signal pulses, which are then "stopped" in the form of the
single atomic coherence $\sigma_{bc}$, the form of which is given
by the third of Eqs (5). Of course the absolute value of this
coherence depends on the place inside the sample but the phase
relations are more general. In Fig.\ref{fig5} we show the argument
of the complex coherence as a function of the phase shift of the
field 3; the simultaneous adiabatic switch-off of the two control
fields has been modeled by a hyperbolic tangent. One can see that
the predictions of the last of Eqs (5) are also excellent.

In Fig. \ref{fig6} we show the space distribution of the coherence
due to the trapped pulse. The length of the sample was $3.5\times
10^{8}$ a.u. and the control pulses were switched off again
simultaneously as a hyperbolic tangent, but somewhat later than
before so that the whole pulse could be trapped inside the medium
without any additional effects of the boundary of the sample. The
distribution can be modeled  with a good accuracy by a sine
square, i.e. the common shape of the incoming pulses. No
parameters have been fitted: the amplitude has been obtained from
the last of Eqs (5), the width is the original width multiplied by
the compression factor $\cos^{2}\theta_{0}$ (the pulse has been
compressed when entering the sample but did not change its width
any more during the storing stage) and the position of the maximum
has been calculated as explained above following Eqs (5).
\section{Conclusions}
We have discussed a simultaneous propagation of two signal pulses
in the medium of four-level atoms in the double $\Lambda$
configuration. We have shown quantitatively that it is practically
possible to divide the incoming pulses into a dark- and
bright-state polaritons, of which the dark one survives during the
evolution. This is connected with the pulses' self-adjusting
during the initial stage of the propagation, so that the condition
of adiabaticity becomes fulfilled. A knowledge of the shape and
position of the dark-state polariton allows one to predict the
characteristics of the transmitted pulses or, in case the pulses
were stored, the value and space distribution of the induced
atomic coherence. Those characteristics can be dynamically
controlled by changing the parameters of the control fields, e.g.
their phase difference. We have also demonstrated that an
adiabatic character of the the pulses' evolution is reflected in a
cancellation of singularities of the atomic susceptibility.

\newpage
\acknowledgments{This work is a part of a program of the National
Laboratory of AMO Physics in Toru\'n, Poland}. 
\newpage
\noindent

\begin{figure}
\caption{Level and coupling schemes; the indices 1 and 3 refer to
signal fields and 2 and 4 - to control fields.} \label{fig1}
\end{figure}
\begin{figure}
\caption{The shape of the propagating initially rectangular pulse
as a function of the local time $t'=t-\frac{z}{c}$, in the case of
a single $\Lambda$ system for a control field of a
time-independent amplitude $\epsilon_{2}=1.2\times10^{-9}$ a.u.:
at the beginning of the sample (1), at the depths: $3\times10^{6}$
a.u. (2), $1.5\times10^{7}$ a.u. (3), $3\times10^{7}$ a.u. (4).
Note the process of smoothing the pulse.} \label{fig2}
\end{figure}
\begin{figure}
\caption{The shapes of the transmitted pulses 1 (curve 1) and 3
(curve 3), compared with their initial shapes (curves 10 and 30,
respectively, for the control fields
$\epsilon_{2}=1.2\times10^{-9}$ a.u. and
$\epsilon_{4}=1.8\times10^{-9}$ a.u. The horizontal lines show the
predicted heights of the transmitted pulses. Note also the
slowdown of the transmitted pulses.} \label{fig3}
\end{figure}
\begin{figure} 
\caption{The shapes of the transmitted pulses 1 (curves $1a$ and
$1b$ and 3 (curves $3a$ and $3b$) for the data as in Fig. 3 but
with additional phase shifts $\phi_{j}$ of the pulse $j$:
$\phi_{2}=\frac{\pi}{2}$, $\phi_{3}=\frac{2\pi}{3}$,
$\phi_{4}=\frac{7\pi}{6}$ (case $a$) and $\phi_{4}=\frac{\pi}{6}$
(case $b$). } \label{fig4}
\end{figure}
\begin{figure}
\caption{The phase shift of the coherence $\sigma_{bc}$ due to the
trapped pulses as a function of the phase shift of the field 3.
The crosses are the results of the numerical computations and the
solid line is obtained from the last of Eqs (5). The control
fields of the maximum values as in Fig. 3 were switched off
simultaneously as a hyperbolic tangent.} \label{fig5}
\end{figure}
\begin{figure}
\caption{The spatial distribution of the coherence $\sigma_{bc}$
due to the trapped pulse.  The crosses are the results of the
numerical computations and the solid line is a sine square, of
which the amplitude has been calculated from the last of Eqs (5),
the position of the maximum is given by the formula following Eqs
(5) and the width is equal to the width of the incoming pulses
divided by $\cos^{2}\theta_{0}$. The control fields were of the
same shape as in Fig. 5 but were switched off somewhat later.}
\label{fig6}
\end{figure}


\newpage
\begin{references}
\bibitem{a1}
S. E. Harris, Phys. Today {\bf 507}, 36 (1997).
\bibitem{a2}
M. O. Scully and M. S. Zubairy, {\em Quantum Optics}, (Cambridge
University Press, 1997).
\bibitem{a3}
C. Liu, Z. Dutton, C. H. Behroozi and L. V. Hau, Nature {\bf 409},
490 (2001).
\bibitem{a4}
D. F. Phillips, A. Fleischhauer, A. Mair, R. L. Walsworth and M.
D. Lukin, Phys. Rev. Lett. {\bf 86}, 783 (2001).
\bibitem{a5}
A. B. Matsko, O. Kocharovskaya, Y. Rostovtsev, G. R. Welch, A. S.
Zibrov, and M. O. Scully, Adv. in At. Mol. Opt. Phys. {\bf 46},
191 (2001).
\bibitem{a6}
M. Fleischhauer and M. D. Lukin, Phys. Rev. Lett. {\bf 84}, 5094
(2000).
\bibitem{a7}
O. Kocharovskaya, Y. Rostovtsev and M. O. Scully, Phys. Rev. Lett.
{\bf 86}, 628 (2001).
\bibitem{a8}
M. D. Lukin, P. R. Hemmer and M. O. Scully, Adv. At. Mol. Opt.
Phys. {\bf 42}, 347 (2000).
\bibitem{a9}
A. Raczy\'nski and J. Zaremba, Opt.Commun. {\bf 209}, 149 (2002).
\end{references}
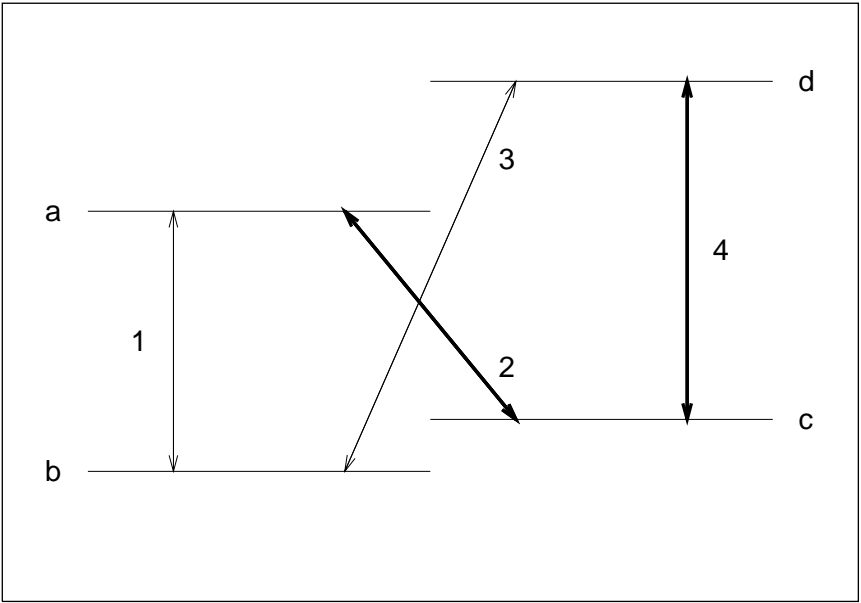
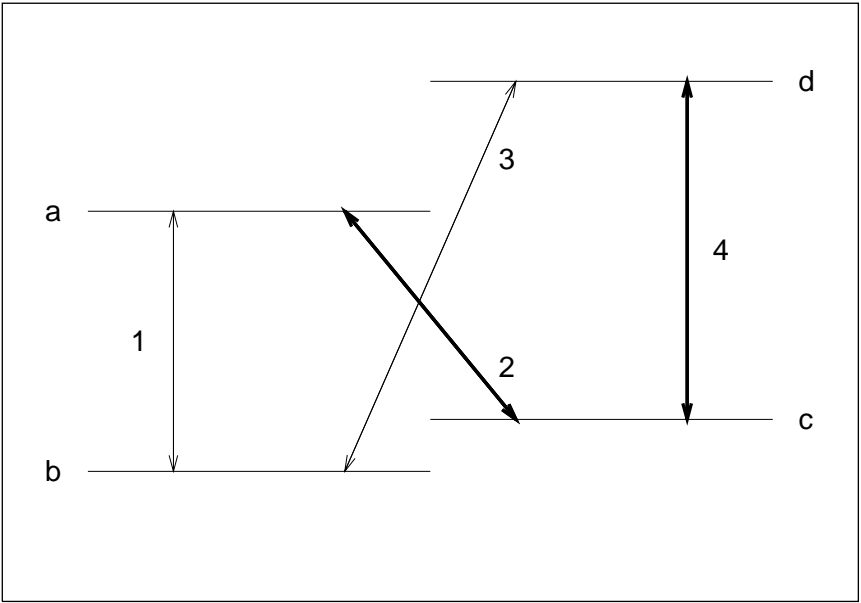
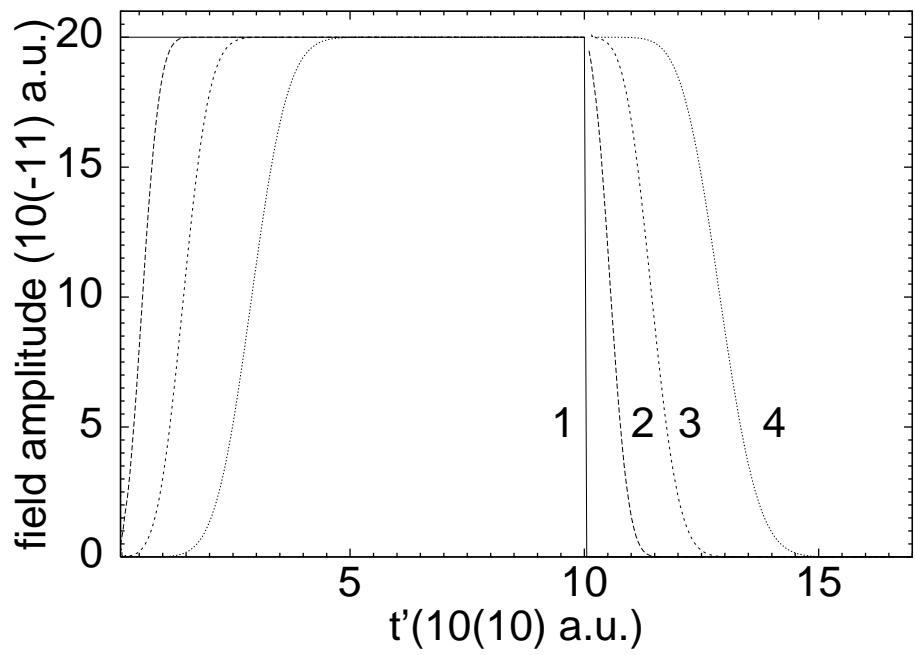
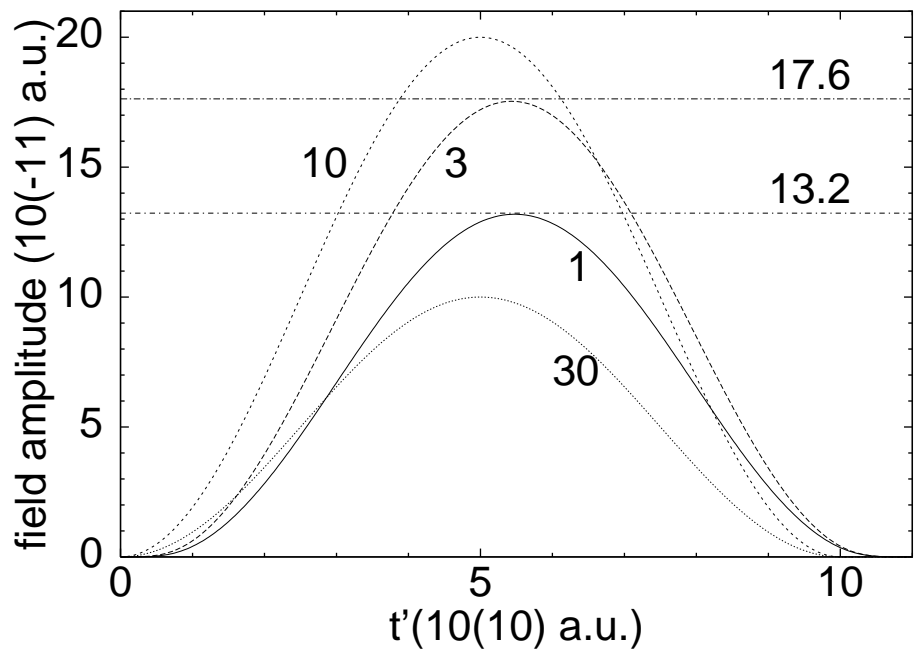
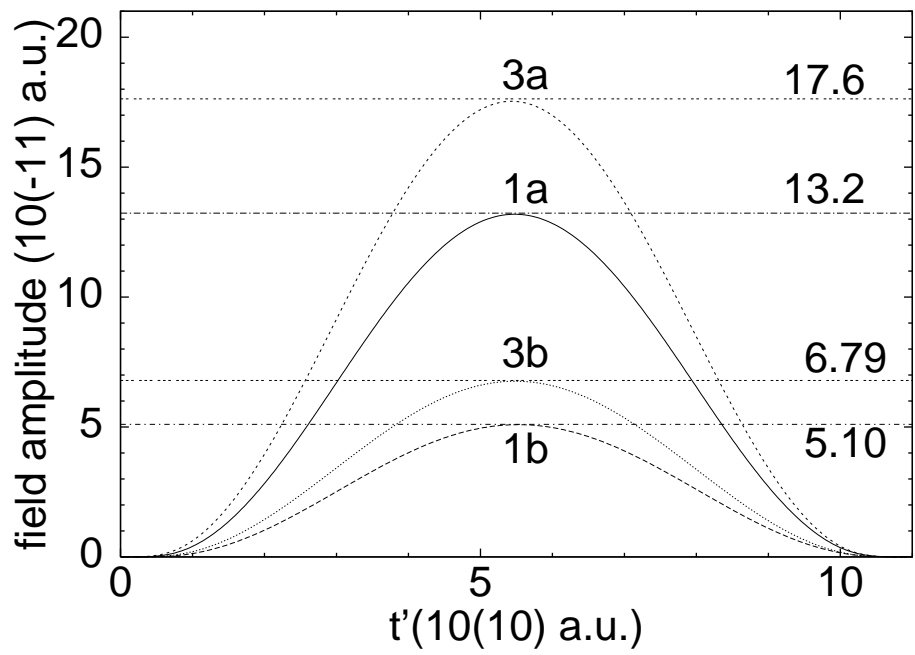
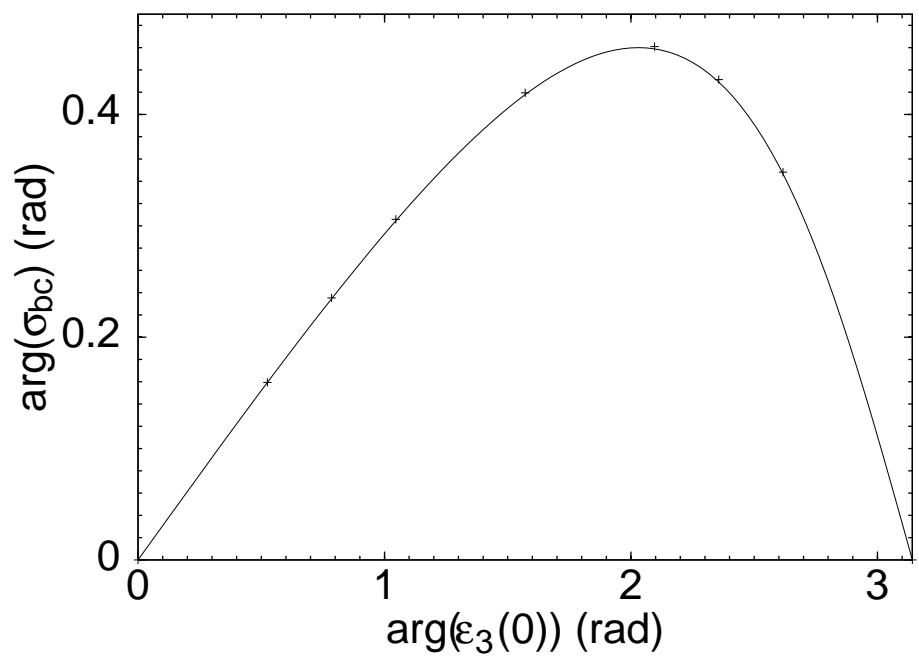
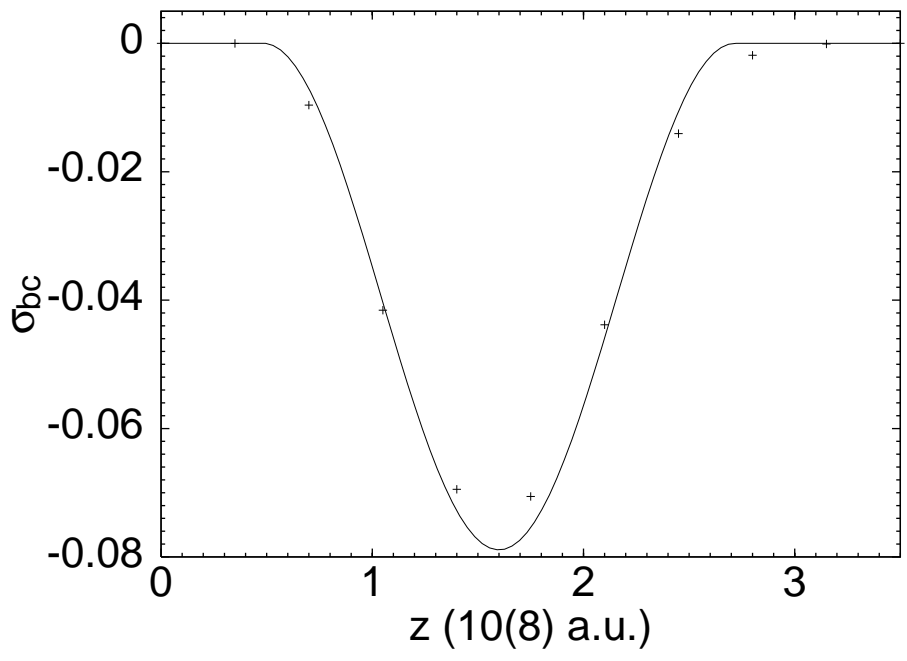
\end{document}